\documentclass[twocolumn, tighten, times, twocolappendix]{aastex631}

\newcommand{\Msun}{M_{\odot}}

\newcommand{\krot}{\kappa_{\rm rot}}

\shorttitle{Halo Angular Momentum Scaling Relations}
\shortauthors{Fall \& Rodriguez-Gomez}

\graphicspath{{./}{figs/}}

\begin{document}

\title{Negligible Effects of Baryons on the Angular Momentum Scaling Relations of Galactic Dark Matter Halos}

\correspondingauthor{S. Michael Fall}
\email{fall@jhu.edu}

\author[0000-0003-3323-9061]{S. Michael Fall}
\affiliation{Department of Physics and Astronomy, Johns Hopkins University, 3400 N. Charles Street, Baltimore, MD 21218, USA}

\author[0000-0002-9495-0079]{Vicente Rodriguez-Gomez}
\affiliation{Instituto de Radioastronom\'ia y Astrof\'isica, Universidad Nacional Aut\'onoma de M\'exico, Apdo. Postal 72-3, 58089 Morelia, Mexico}


\begin{abstract}
In cosmological simulations without baryons, the relation between the specific angular momentum $j_{\rm h}$ and mass $M_{\rm h}$ of galactic dark matter halos has the well-established form $j_{\rm h} \propto M_{\rm h}^{2/3}$. This is invariably adopted as the starting point in efforts to understand the analogous relation between the specific angular momentum $j_{\ast}$ and mass $M_{\ast}$ of the stellar parts of galaxies, which are often re-expressed relative to the corresponding halo properties through the retention fractions $f_j = j_{\ast} / j_{\rm h}$ and $f_M = M_{\ast} / M_{\rm h}$. An important caveat here is that the adopted $j_{\rm h} \propto M_{\rm h}^{2/3}$ relation could, in principle, be modified by the gravitational back-reaction of baryons on dark matter (DM). We have tested for this possibility by comparing the $j_{\rm h}$--$M_{\rm h}$ relations in the IllustrisTNG100 and TNG50 simulations that include baryons (full-physics runs) with their counterparts that do not (DM-only runs). In all cases, we find scaling relations of the form $j_{\rm h} \propto M_{\rm h}^{\alpha}$, with $\alpha \approx 2/3$ over the ranges of mass and redshift studied here: $M_{\rm h} \geq 10^{10} \, M_{\odot}$ and $0 \leq z \leq 2$. The values of $\alpha$ are virtually identical in the full-physics and DM-only runs at the same redshift. The only detectable effect of baryons on the $j_{\rm h}$--$M_{\rm h}$ relation is a slightly higher normalization, by 12\%--15\% at $z=0$ and by 5\% at $z=2$. This implies that existing estimates of $f_j$ based on DM-only simulations should be adjusted downward by similar amounts. Finally, we discuss briefly some implications of this work for studies of galaxy formation.
\end{abstract}


\section{Introduction}
\label{sec:intro}

Two of the most basic properties of cosmic structures are their mass $M$ and angular momentum $J$ or, equivalently, $M$ and specific angular momentum $j = J/M$. The relation between $j$ and $M$ for a population of objects reflects the physical processes by which they form and evolve. In cold dark matter-type cosmologies, bound structures in virial equilibrium develop from small perturbations in the early universe by gravitational instability. Throughout this development, but especially in the translinear phase of growth, they acquire angular momentum from the tidal torques exerted by neighboring perturbations \citep{Peebles1969, Doroshkevich1970, White1984}.

These processes are now well understood in cosmologies that include only gravitationally interacting dark matter (DM), usually thought to provide a good description of galactic halos. The results of cosmological $N$-body simulations are often expressed in terms of the spin parameter $\lambda \equiv j |E|^{1/2} / (G M^{3/2})$, where $E$ is the total energy (potential plus kinetic) of a halo, and $G$ is the gravitational constant. The median spin value, derived from many DM-only (DMO) simulations, is $\hat{\lambda} \approx 0.035$, irrespective of cosmological parameters and the mass and density contrast of the halos \citep{Bullock2001, vandenBosch2002, Avila-Reese2005, Bett2007, Maccio2007, Maccio2008, Zjupa2017}. This implies a relation between the specific angular momentum $j$, mass $M$, and mean internal density $\left<\rho\right>$ of halos of the form $j \propto \left<\rho\right>^{-1/6} M^{2/3}$ (given the scalings $E \propto M^2 / R$ and $\left<\rho\right> \propto M / R^3$ with mass $M$ and radius $R$).  For a fixed density contrast, $\left<\rho\right>/\rho_{\rm crit} =$ constant, this becomes $j \propto M^{2/3}$.  The same simulations show that the dispersions of $\lambda$ and $j$ about their median values at each $M$ are substantial: $\sigma (\ln\lambda) \approx \sigma (\ln j) \approx 0.5$--$0.6$.

The optically visible stellar components of galaxies obey a remarkably similar scaling relation: $j = A M^{\alpha}$, with an exponent $\alpha \approx 0.6$ and an amplitude $A$ that correlates with disk fraction and morphological type \citep{Fall1983, Romanowsky2012, Fall2013, Obreschkow2014, Fall2018, Posti2018a, DiTeodoro2021, ManceraPina2021, ManceraPina2021a, Hardwick2022, DiTeodoro2023, Pulsoni2023}. This apparently simple stellar $j$--$M$ relation is less straightforward to interpret than the corresponding halo relation because it must reflect the complicated combined effects of cooling, collapse, star formation, black hole growth, and feedback in the baryonic components of galaxies. With this in mind, it is often useful to re-express the stellar $j$--$M$ relation relative to the better-understood halo relation in terms of the ``retention fractions'' for mass, $f_M \equiv M_{\ast} / M_{\rm h}$, and specific angular momentum, $f_j \equiv j_{\ast} / j_{\rm h}$, the subscripts $\ast$ and h now specifying stellar and halo quantities explicitly. The mass retention fraction $f_M$ is also known as the stellar-to-halo mass relation (SHMR).

In calculations of the retention fraction for specific angular momentum $f_j$, the reference $j$--$M$ relation is invariably taken to be the one derived from DMO simulations, $j_{\rm h} \propto M_{\rm h}^{2/3}$. This is equivalent to assuming, at least implicitly, that the baryonic processes involved in the formation of the luminous bodies of galaxies have a negligible effect on the halo $j$--$M$ relation. While plausible, this is not guaranteed to be true. Indeed, \cite{Du2022} have suggested that the gravitational back-reaction of baryons on the halo $j$--$M$ relation is significant and necessary to explain the observed stellar $j$--$M$ relation and the associated retention fraction $f_j$. We tested for this possibility during our recent comprehensive study of galactic angular momentum in the IllustrisTNG simulations \citep{Rodriguez-Gomez2022} and reached a different conclusion. The purpose of this Letter is to present the results of these tests.

\section{Simulations and Analysis}
\label{sec:methods}

We examine the $j$--$M$ and $\lambda$--$M$ relations of halos in four simulations of the IllustrisTNG suite: the TNG100 and TNG50 runs, which include both baryons and DM, and their DMO counterparts, TNG100-Dark and TNG50-Dark. Our analysis is based on quantities we have computed from the positions and velocities of particles at different snapshots in the publicly accessible IllustrisTNG data files \citep{Nelson2019}. For complete descriptions of the simulations, we refer interested readers to the original papers \citep{Marinacci2018, Naiman2018, Nelson2018, Pillepich2018a, Springel2018, Nelson2019a, Pillepich2019}. Here, we provide only a brief summary, sufficient to interpret the results presented in Section \ref{sec:results}.

The TNG100 and TNG50 runs include plausible subgrid prescriptions for star formation and black hole growth, as well as feedback from supernovae and active galactic nuclei (AGNs). We refer to them as ``full-physics'' (FP) runs.  These simulations produce populations of galaxies that resemble real ones in many respects, including galactic masses and specific angular momenta \citep{Du2022, Rodriguez-Gomez2022}. The main differences between the TNG100 and TNG50 runs are their volumes (cubes with comoving $\sim$100 Mpc and $\sim$50 Mpc sides) and numerical resolutions ($7.5 \times 10^6 \, \Msun$ for DM and $1.4 \times 10^6 \, \Msun$ for baryons in TNG100, and 16 times better for both in TNG50). Thus, TNG100 is preferred for studies of massive galaxy populations and TNG50 for low-mass populations. The DMO runs have the same volumes and initial conditions as the corresponding FP runs. To make precise comparisons between the simulations with and without baryons, we have matched the halos in the DMO runs individually with their counterparts in the FP runs using the procedure described by \cite{Rodriguez-Gomez2017}.

Halos in the simulations are identified by the friends-of-friends (FoF) algorithm \citep[][]{Davis1985} and gravitationally bound subhalos within them by the \textsc{subfind} algorithm \citep{Springel2001, Dolag2009a}. Following standard practice, we measure positions within each halo relative to the most tightly bound particle and velocities relative to the center-of-mass motion\footnote{The specific angular momentum computed in this reference frame is exactly the same as that in a frame defined by both the center-of-mass position and velocity, as one can show by simple algebra.} \citep[][]{Genel2015, Zjupa2017}. We compute all halo quantities, denoted by the subscript h, by including both DM and baryons within $R_{\rm h} \equiv R_{\rm 200, crit}$, the radius enclosing the mean density $\left<\rho\right> = 200 \rho_{\rm crit}$ (aka the ``virial radius''); hence, $M_{\rm h} \equiv M_{\rm 200,crit}$, $J_{\rm h} \equiv J_{\rm 200,crit}$, and $j_{\rm h} \equiv J_{\rm h} / M_{\rm h}$.\footnote{Note that, by definition, halo quantities such as $M_{\rm h}$ and $j_{\rm h}$ include all material within $R_{\rm h}$ in both the central and satellite subhalos as well as any unbound particles (aka ``fuzz'').} As an approximation to the spin parameter $\lambda$, we compute the analogous quantity $\lambda^{\prime} \equiv j_{\rm h} / (\sqrt{2} R_{\rm h} V_{\rm h})$, where $V_{\rm h} = (G M_{\rm h} / R_{\rm h})^{1/2}$ is the circular velocity at the virial radius \citep{Bullock2001}. For a singular isothermal halo, $\lambda^{\prime}$ is exactly the same as $\lambda$ (in the limit $\lambda \ll 1$).

Galaxies in the simulations consist of all the star particles and gas elements within the subhalos identified by \textsc{subfind}. For each FoF halo, this algorithm assigns a smooth central or ``background'' subhalo by removing all its satellite subhalos while checking that it remains gravitationally bound \citep[][Figure 3]{Springel2001}. Thus, central galaxies---which we consider exclusively throughout this work---consist of all the baryonic material associated with these central subhalos. We specify galactic morphology in terms of the parameter $\kappa_{\rm rot}$, defined as the fraction of the stellar kinetic energy invested in rotation \citep{Sales2010, Rodriguez-Gomez2017}. In particular, we classify galaxies with $\kappa_{\rm rot} < 0.5$ as spheroid dominated and those with $\kappa_{\rm rot} \geq 0.5$ as disk dominated. We compute the stellar properties of galaxies, such as $M_{\ast}$ and $\kappa_{\rm rot}$, by summing over all the stars bound to their subhalos.

\section{Results}
\label{sec:results}

\begin{figure*}
  \centerline{\hbox{
  	\includegraphics[width=17.5cm]{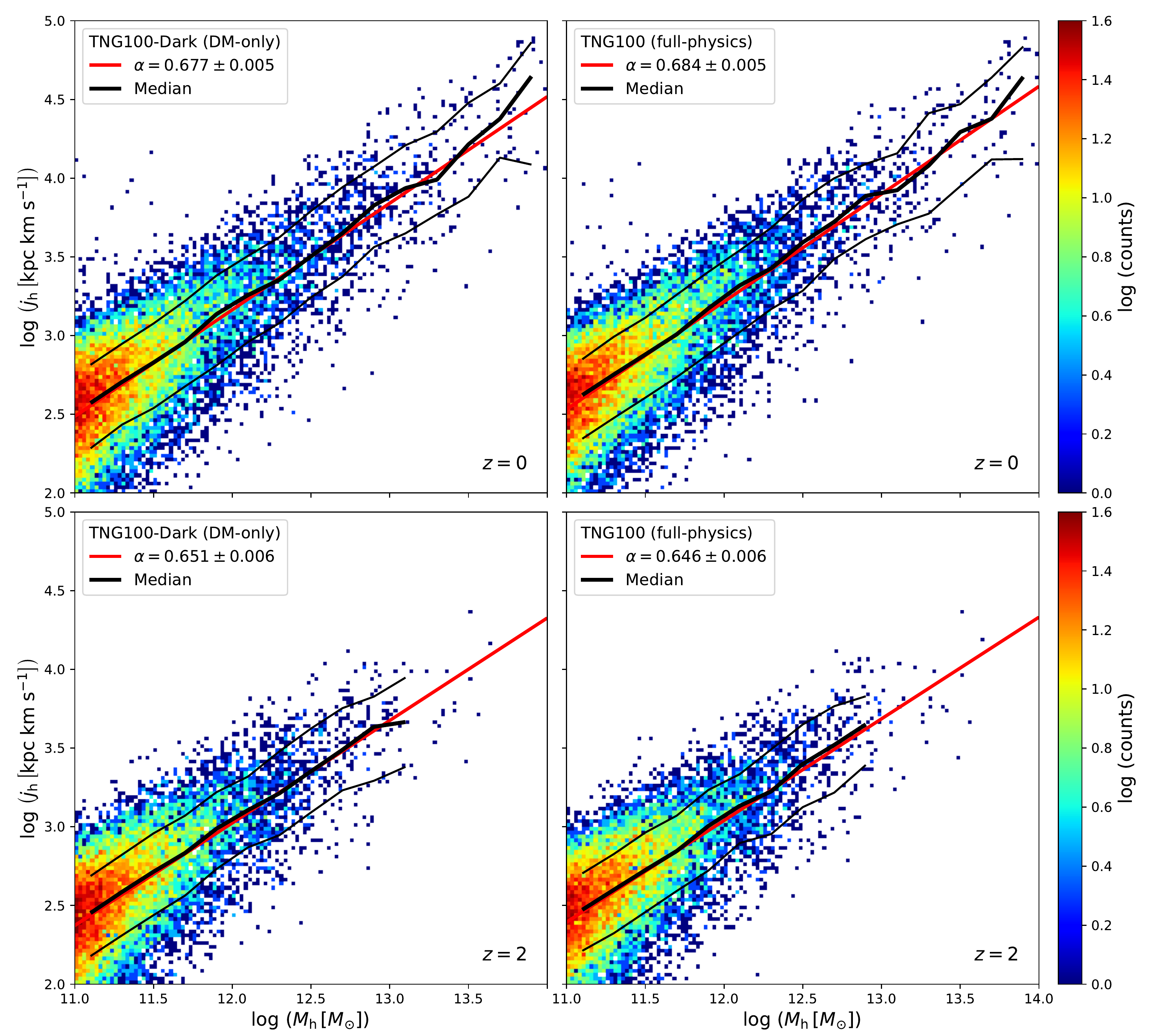}
  }}
	\caption{Specific angular momentum $j_{\rm h}$ plotted against mass $M_{\rm h}$ for halos in the TNG100 DMO and FP runs (left and right panels, respectively) at redshifts $z = 0$ and $z = 2$ (upper and lower panels, respectively). The color scale indicates the number of halos in each 2D bin. The red lines represent the best-fit linear relations, while the black lines represent the running median relations and 16th--84th percentile ranges. Note the close similarity of the halo $j$--$M$ relations in the DMO and FP runs and the slightly flatter slopes at $z = 2$ than at $z = 0$.}
	\label{fig:j200_vs_m200_TNG100_many_z}
\end{figure*}

\begin{figure*}
  \centerline{\hbox{
  	\includegraphics[width=17.5cm]{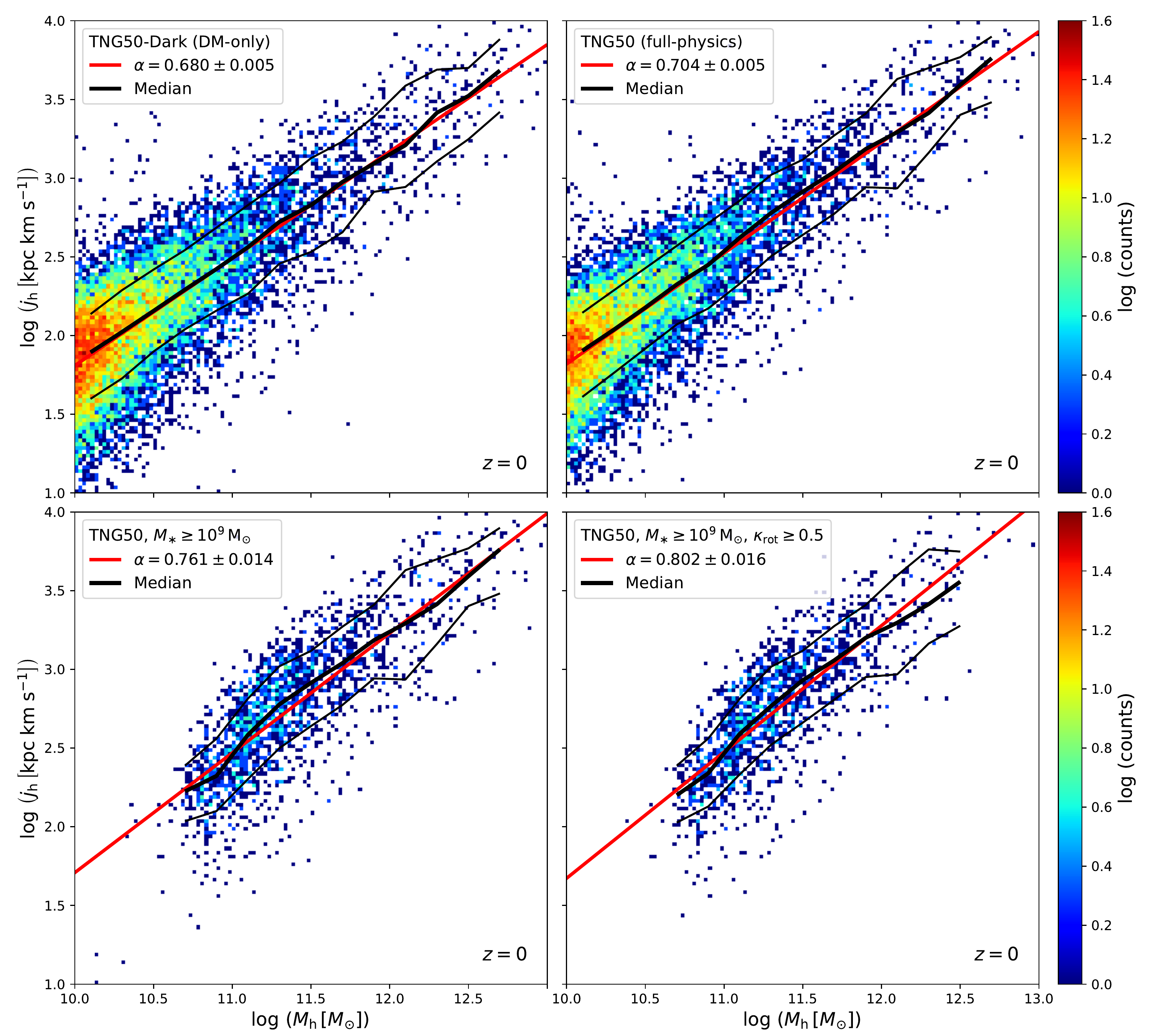}
  }}
	\caption{
Specific angular momentum $j_{\rm h}$ plotted against mass $M_{\rm h}$ for halos in the TNG50 DMO and FP runs at $z = 0$ (upper left and upper right panels, respectively) and the FP run at $z = 0$ with selection criteria based on the stellar masses and morphologies of galaxies: $M_{\ast} \geq 10^{9} \, \Msun$ (lower left panel) and $M_{\ast} \geq 10^{9} \, \Msun$ and $\krot \geq 0.5$ (lower right panel). The last of these replicates the selection criteria adopted by \protect\cite{Du2022}. The color scale indicates the number of halos in each 2D bin. The red lines represent the best-fit linear relations, while the black lines represent the running median relations and 16th--84th percentile ranges. Note the close similarity of the halo $j$--$M$ relations in the DMO and FP runs and the steeper fitted slopes in the samples selected by stellar properties.}
	\label{fig:j200_vs_m200_TNG50_selection_effects}
\end{figure*}

\begin{figure}
  \centerline{\hbox{
  	\includegraphics[width=9cm]{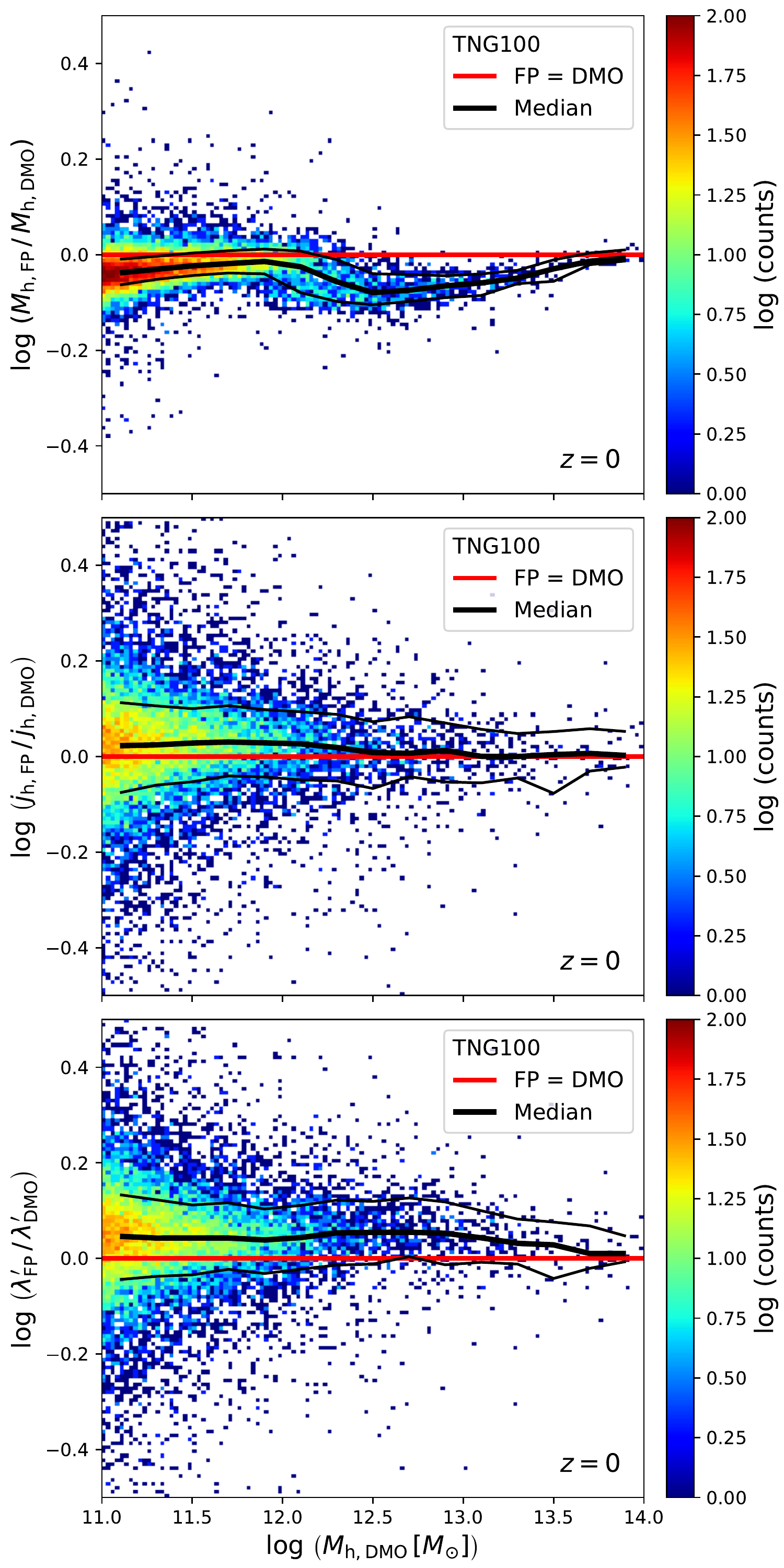}
  }}
	\caption{
Ratios of mass $M_{\rm h}$ (top panel), specific angular momentum $j_{\rm h}$ (middle panel), and spin parameter $\lambda^{\prime}$ (bottom panel) for halos in the FP run to the same quantities for the individually matched halos in the DMO run plotted against mass $M_{\rm h,DMO}$ for TNG100 at $z = 0$. The color scale indicates the number of halos in each 2D bin. The red horizontal lines indicate equality, while the black lines represent the running median relations and 16th--84th percentile ranges. This diagram shows directly the effects of baryons on the halo properties $M_{\rm h}$, $j_{\rm h}$, and $\lambda^{\prime}$. Note the relatively small and nearly constant median offset of $\lambda^{\prime}_{\rm FP}$ from $\lambda^{\prime}_{\rm DMO}$, indicating that the halo relation $j_{\rm h} \propto M_{\rm h}^{\alpha}$ with $\alpha \approx 2/3$ is preserved by baryonic processes.}
	\label{fig:hydro_to_dmonly_3panels}
\end{figure}

The main results of this Letter---comparisons between the halo $j$--$M$ and $\lambda^{\prime}$--$M$ relations in the FP and DMO IllustrisTNG simulations---are displayed in Figures \ref{fig:j200_vs_m200_TNG100_many_z}--\ref{fig:hydro_to_dmonly_3panels} and Table \ref{tab:best-fit_values}. For these comparisons, we have performed linear regressions of the form

\begin{equation}
\log j_{\rm h} = \alpha \log\left(\frac{M_{\rm h}}{10^{12} \, \Msun}\right) + \log j_{\rm h, 12}
\label{eq:j200_vs_M200}
\end{equation}

\begin{equation}
\log \lambda^{\prime} = \beta \log\left(\frac{M_{\rm h}}{10^{12} \, \Msun}\right) + \log \lambda^{\prime}_{12}
\label{eq:lambda_vs_M200}
\end{equation}
to determine the best-fit slopes $\alpha$ and $\beta$ and normalizations $\log j_{\rm h, 12}$ and $\log \lambda^{\prime}_{12}$. We then compute the standard (i.e., $1 \sigma$) errors of these quantities by bootstrap resampling. The figures and table present results for two redshifts, $z = 0$ and $z = 2$, and several different selection criteria specified by lower limits on halo mass $M_{\rm h}$, stellar mass $M_{\ast}$, and morphology parameter $\krot$.

We have adopted these redshifts and selection criteria with the following thoughts in mind. Because the masses of individual halos grow with time, so also does the upper extent of their $j$--$M$ relation (compare the upper and lower panels of Figure \ref{fig:j200_vs_m200_TNG100_many_z}). As a result, the available range of masses shrinks with increasing redshift, and above $z \sim 2$, it becomes too narrow to determine the slopes $\alpha$ and $\beta$ reliably. The restrictions on halo masses, $M_{\rm h} \geq 10^{11} \, \Msun$ for TNG100 and $M_{\rm h} \geq 10^{10} \, \Msun$ for TNG50, were chosen to compensate approximately for the different resolutions and volumes of these simulations. These limits are also intended to exclude the low-mass galaxies that might suffer from the spurious heating of stellar motions by DM particles and the resulting transfer of angular momentum to their halos \citep{Ludlow2021, Wilkinson2023}. We have also checked what happens to the fitted slopes $\alpha$ and $\beta$ when we impose the same selection criteria as \cite{Du2022} on the stellar properties of galaxies, namely $M_{\ast} \geq 10^{9} \, \Msun$ and $\krot \geq 0.5$.

We first compare the halo $j$--$M$ and $\lambda^{\prime}$--$M$ relations in the FP and DMO runs in samples defined by halo properties alone, specifically $M_{\rm h} \geq 10^{11} \, \Msun$ for TNG100 and $M_{\rm h} \geq 10^{10} \, \Msun$ for TNG50. The upper and lower panels of Figure \ref{fig:j200_vs_m200_TNG100_many_z} show the halo $j$--$M$ relation in the TNG100 FP and DMO runs at $z = 0$ and $z = 2$, while the upper panels of Figure \ref{fig:j200_vs_m200_TNG50_selection_effects} show the same results for TNG50 at $z = 0$. Table \ref{tab:best-fit_values} lists the slopes $\alpha$ and $\beta$ of the halo $j$--$M$ and $\lambda^{\prime}$--$M$ relations in all the simulations at $z = 0$ and $z = 2$. In all cases, these are close to the expected values $\alpha = 2/3$ and $\beta = 0$. Furthermore, the slopes in the FP runs are nearly the same as those in the corresponding DMO runs, with $\Delta \alpha \approx \Delta \beta < 0.01$ for TNG100 and $\Delta \alpha \approx \Delta \beta < 0.03$ for TNG50. However, for both TNG100 and TNG50, the normalizations of the halo $j$--$M$ and $\lambda^{\prime}$--$M$ relations are slightly higher in the FP runs than in the DMO runs, by $\Delta \log j_{\rm h,12} \approx \Delta \log \lambda^{\prime}_{\rm 12} \approx 0.05$--$0.06$ (12\%--15\%) at $z = 0$ and $\approx 0.02$ (5\%) at $z = 2$. We also find that these relations are slightly flatter at $z = 2$ than at $z = 0$, by $\Delta \alpha \approx \Delta \beta \approx 0.02$--$0.05$ for both TNG100 and TNG50.

We can elucidate these results by comparing the masses $M_{\rm h}$, specific angular momenta $j_{\rm h}$, and spin parameters $\lambda^{\prime}$ of halos in the FP runs with those of their individually matched counterparts in the DMO runs. These comparisons are shown in Figure \ref{fig:hydro_to_dmonly_3panels}, where we plot the ratios $M_{\rm h,FP} / M_{\rm h,DMO}$, $j_{\rm h,FP} / j_{\rm h,DMO}$, and $\lambda^{\prime}_{\rm FP} / \lambda^{\prime}_{\rm DMO}$ against $M_{\rm h,DMO}$ for the TNG100 simulation at $z = 0$. Evidently, baryons have a rather complex effect on the median offset between $M_{\rm h,FP}$ and $M_{\rm h,DMO}$, likely caused by the varying strengths of supernova and AGN feedback, but they introduce relatively little scatter about this trend (top panel). In contrast, the median offsets between $j_{\rm h, FP}$ and $j_{\rm h, DMO}$ and between $\lambda^{\prime}_{\rm FP}$ and $\lambda^{\prime}_{\rm DMO}$ are small and nearly constant, while the scatter about these trends is large (middle and bottom panels). Since the spin parameter can be re-expressed as $\lambda^{\prime} \propto \left<\rho\right>^{1/6} j_{\rm h} / M_{\rm h}^{2/3} \propto j_{\rm h} / M_{\rm h}^{2/3}$ (at fixed $\left<\rho\right> / \rho_{\rm crit}$), the nearly constant median offset of $\lambda^{\prime}$ with $M_{\rm h}$ ensures that the relation $j_{\rm h} \propto M_{\rm h}^{\alpha}$ with $\alpha \approx 2/3$ is preserved, but with a 12\%--15\% higher amplitude, consistent with the results listed in Table \ref{tab:best-fit_values}.

Why does the inclusion of baryons in the simulations make so little difference to the halo $j$--$M$ relation?  There are three contributing factors, corresponding to the quantities $M$, $j$, and $E$ that appear in the formula for $\lambda$ (and hence $\lambda^{\prime}$). First, the fraction of mass in baryons is small compared with that in DM (16\% vs 84\%). Second, while the baryons within individual halos can gain or lose large amounts of specific angular momentum, there is little net transfer within the population of halos as a whole. Third, the ratios of the specific binding energies of circumgalactic baryons and DM particles within the same halos, as measured by their kinetic temperatures, are of order unity and increase only gradually with mass. These factors help to explain the results shown in Figure \ref{fig:hydro_to_dmonly_3panels}, particularly the near constancy of $\lambda^{\prime}$ with $M_{\rm h}$ and thus $j_{\rm h} \propto M_{\rm h}^{2/3}$.

The preceding results indicate that baryons have negligible \textit{physical} effects on the slope $\alpha$ of the halo $j$--$M$ relation in the IllustrisTNG simulations. This does not mean, however, that this slope will always be the same for samples of galaxies selected by different baryonic properties, such as stellar mass $M_{\ast}$ and morphology parameter $\krot$. In these cases, there can be non-negligible \textit{apparent} effects on $\alpha$. This is illustrated for TNG50 at $z = 0$ in the lower panels of Figure \ref{fig:j200_vs_m200_TNG50_selection_effects}---on the left for the restriction $M_{\ast} \geq 10^{9} \, \Msun$ alone and on the right when combined with the restriction $\krot \geq 0.5$. The latter mimics a luminosity-limited sample of disk-dominated galaxies. In this case, we find $\alpha \approx 0.8$, similar to the slope \cite{Du2022} found. The reason for this apparent increase in $\alpha$ can be traced to the fact that a vertical cut in the $j_{\rm h}$--$M_{\ast}$ plane maps into a diagonal cut in the $j_{\rm h}$--$M_{\rm h}$ plane, as shown in the lower panels of Figure \ref{fig:j200_vs_m200_TNG50_selection_effects} here and in Figure 5 of \cite{Du2022}. Consequently, the fitted slope of the halo $j$--$M$ relation for this particular sample is slightly steeper than that for the underlying population.

\section{Discussion}
\label{sec:discussion}

\begin{table*}
\begin{center}
\caption{Regression Fits to Equations (\ref{eq:j200_vs_M200}) and (\ref{eq:lambda_vs_M200}) for Different Simulations, Redshifts, and Selection Criteria.}
\label{tab:best-fit_values}
\begin{tabular}{ l c c c c r c }

\hline
\hline

Simulation & Redshift & Selection criteria & $\alpha$ & $\log j_{\rm h, 12}$ & \multicolumn{1}{c}{$\beta$} & $\log \lambda^{\prime}_{12}$ \\

(1) & (2) & (3) & (4) & (5) & \multicolumn{1}{c}{(6)} & (7) \\

\hline

\vspace{-1.2em} \\

\textbf{TNG100} & $\bf z = 0$ \\

DM-only &  & $M_{\rm h} \geq 10^{11} \, \Msun$ & $0.677 \pm 0.005$ & $3.164 \pm 0.003$ & $0.010 \pm 0.005$ & $-1.464 \pm 0.003$ \\

Full-physics &  & $M_{\rm h} \geq 10^{11} \, \Msun$ & $0.684 \pm 0.005$ & $3.215 \pm 0.003$ & $0.017 \pm 0.005$ & $-1.412 \pm 0.003$ \\

Full-physics &  & $M_{\ast} \geq 10^{9} \, \Msun$ & $0.746 \pm 0.005$ & $3.212 \pm 0.003$ & $0.080 \pm 0.005$ & $-1.416 \pm 0.003$ \\

Full-physics &  & $M_{\ast} \geq 10^{9} \, \Msun$, $\kappa_{\rm rot} \geq 0.5$ & $0.781 \pm 0.007$ & $3.244 \pm 0.004$ & $0.114 \pm 0.007$ & $-1.384 \pm 0.004$ \\

\vspace{-1.2em} \\

\textbf{TNG50} \\

DM-only &  & $M_{\rm h} \geq 10^{10} \, \Msun$ & $0.680 \pm 0.005$ & $3.169 \pm 0.008$ & $0.013 \pm 0.005$ & $-1.459 \pm 0.008$ \\

Full-physics &  & $M_{\rm h} \geq 10^{10} \, \Msun$ & $0.704 \pm 0.005$ & $3.227 \pm 0.008$ & $0.038 \pm 0.005$ & $-1.399 \pm 0.008$ \\

Full-physics &  & $M_{\ast} \geq 10^{9} \, \Msun$ & $0.761 \pm 0.014$ & $3.230 \pm 0.010$ & $0.094 \pm 0.014$ & $-1.397 \pm 0.010$ \\

Full-physics &  & $M_{\ast} \geq 10^{9} \, \Msun$, $\kappa_{\rm rot} \geq 0.5$ & $0.802 \pm 0.016$ & $3.278 \pm 0.013$ & $0.136 \pm 0.016$ & $-1.349 \pm 0.013$ \\

\hline

\vspace{-1.2em} \\

\textbf{TNG100} &  $\bf z = 2$ \\

DM-only &  & $M_{\rm h} \geq 10^{11} \, \Msun$ & $0.651 \pm 0.006$ & $3.024 \pm 0.004$ & $-0.014 \pm 0.006$ & $-1.431 \pm 0.004$ \\

Full-physics &  & $M_{\rm h} \geq 10^{11} \, \Msun$ & $0.646 \pm 0.006$ & $3.040 \pm 0.004$ & $-0.018 \pm 0.006$ & $-1.409 \pm 0.004$ \\

Full-physics &  & $M_{\ast} \geq 10^{9} \, \Msun$ & $0.746 \pm 0.007$ & $3.044 \pm 0.004$ & $0.080 \pm 0.007$ & $-1.405 \pm 0.004$ \\

Full-physics &  & $M_{\ast} \geq 10^{9} \, \Msun$, $\kappa_{\rm rot} \geq 0.5$ & $0.730 \pm 0.012$ & $3.017 \pm 0.006$ & $0.064 \pm 0.012$ & $-1.431 \pm 0.006$ \\

\vspace{-1.2em} \\

\textbf{TNG50} \\

DM-only &  & $M_{\rm h} \geq 10^{10} \, \Msun$ & $0.659 \pm 0.005$ & $3.026 \pm 0.008$ & $-0.006 \pm 0.005$ & $-1.427 \pm 0.008$ \\

Full-physics &  & $M_{\rm h} \geq 10^{10} \, \Msun$ & $0.650 \pm 0.005$ & $3.042 \pm 0.008$ & $-0.013 \pm 0.005$ & $-1.403 \pm 0.008$ \\

Full-physics &  & $M_{\ast} \geq 10^{9} \, \Msun$ & $0.722 \pm 0.020$ & $3.031 \pm 0.013$ & $0.056 \pm 0.020$ & $-1.415 \pm 0.013$ \\

Full-physics &  & $M_{\ast} \geq 10^{9} \, \Msun$, $\kappa_{\rm rot} \geq 0.5$ & $0.683 \pm 0.030$ & $2.983 \pm 0.017$ & $0.016 \pm 0.030$ & $-1.463 \pm 0.017$ \\

\hline

\end{tabular}
\end{center}
\tablecomments{The quoted $1 \sigma$ errors were derived by bootstrap resampling.}
\end{table*}

Our main conclusion is that the halo $j$--$M$ and $\lambda^{\prime}$--$M$ relations in the IllustrisTNG simulations are very close to the expected forms, $j_{\rm h} \propto M_{\rm h}^{2/3}$ and $\lambda^{\prime} \propto M_{\rm h}^0$, at least over the ranges of mass and redshift examined here: $M_{\rm h} \geq 10^{10} \, \Msun$ and $0 \leq z \leq 2$. This is true for both TNG100 and TNG50 and for both FP and DMO runs. We find some deviations from the canonical slopes $\alpha = 2/3$ and $\beta = 0$ that are statistically significant given the exceedingly small formal errors listed in Table \ref{tab:best-fit_values}. However, these small deviations are comparable to the differences in $\alpha$ and $\beta$ between the TNG100 and TNG50 simulations and are likely negligible for all practical purposes. In particular, the assumed halo $j$--$M$ relation with $\alpha = 2/3$ is sufficiently accurate to estimate the retention fractions of specific angular momentum, $f_j = j_{\ast} / j_{\rm h}$, from the observed stellar $j$--$M$ relations for low-redshift galaxies. Beyond $z \approx 2$, the reference value of $\alpha$ may be slightly lower than $2/3$.

The halo $j$--$M$ relation---in the form confirmed here---when combined with the observed stellar $j$--$M$ relation, provides an important link between the retention fractions $f_M$ and $f_j$, and with it some valuable insights into galaxy formation, as we now summarize briefly. For the power-law models, $j_{\ast} = A_{\ast} M_{\ast}^{\alpha_{\ast}}$ and $j_{\rm h} = A_{\rm h} M_{\rm h}^{\alpha_{\rm h}}$, we have
\begin{equation}
f_j / f_M^{\alpha_{\rm h}} = \left(A_{\ast} / A_{\rm h}\right) M_{\ast}^{\alpha_{\ast} – \alpha_{\rm h}} \sim {\rm constant}.
\label{eq:fj_vs_fM}
\end{equation}
Given the small difference between the exponents of the stellar and halo $j$--$M$ relations ($\alpha_{\ast} \approx 0.6$ vs $\alpha_{\rm h} = 2/3$), Equation (\ref{eq:fj_vs_fM}) indicates that $f_j$ and $f_M$ will have similar shapes although the former will be subdued relative to the latter. This is interesting because recent dynamical studies have revealed that the high-mass shape of the SHMR depends strongly on galactic morphology \citep{Posti2019, Posti2019a, Posti2021, DiTeodoro2023}. For disk-dominated galaxies, $f_M$ rises monotonically with mass, with no prominent features, while for spheroid-dominated galaxies, $f_M$ has the more familiar, inverted-U shape, with a peak near the mass of the Milky Way. Another simple consequence of Equation (\ref{eq:fj_vs_fM}) is that the value of $f_j$ for each type of galaxy scales directly with the stellar amplitude $A_{\ast}$ and inversely with the halo amplitude $A_{\rm h}$.

Empirical determinations of the stellar $j$--$M$ relation require both photometric and kinematic data, ideally covering each galaxy in a large sample in two dimensions to large radii. For spiral galaxies, the stellar $j$--$M$ relation is quite secure; all determinations, including the original one 40 yr ago, are in remarkably close agreement. This is a consequence of the known inclination of each galactic disk and the fact that the angular momentum of disks with exponential surface-density profiles and flat rotation curves converges rapidly beyond about two effective radii. Recent studies of the stellar $j$--$M$ relation of low-redshift spiral galaxies all find very similar angular momentum retention fractions, $f_j \approx 0.8$, based on $A_{\rm h}$ values from DMO simulations, with little or no dependence on mass \citep{Fall2013, Fall2018, Posti2018a, Posti2019a, DiTeodoro2021, DiTeodoro2023}. Adjusting this for the higher $A_{\rm h}$ values in simulations with baryons gives $f_j \approx 0.7$. A corollary of this result is that, on average, the exponential scale radii $R_{\rm d}$ of galactic disks are related to the virial radii $R_{\rm h}$ of their halos by $R_{\rm d} / R_{\rm h} \approx f_j \hat{\lambda} / \sqrt{2} \sim 0.02$ (\citealt{Fall1980, Fall1983, Mo1998}; note the changes in notation, $\alpha \rightarrow 1 / R_{\rm d}$ and $r_{\rm t} \rightarrow R_{\rm h}$, between the early and later papers and the fact that the product $f_j \hat{\lambda}$ is independent of the $A_{\rm h}$ adjustment).

Determinations of the stellar $j$--$M$ relations of lenticular and elliptical galaxies are much harder because the inclinations of the galaxies are uncertain, the shapes of the rotation curves vary, the angular momenta converge slowly, and kinematic data are sparse at large radii. The resulting estimates of the angular momentum retention fractions of spheroid-dominated galaxies are $f_j \sim 0.1$, with an uncertain mass dependence \citep{Romanowsky2012, Fall2013, Fall2018, Pulsoni2023}. The observed order-of-magnitude difference between the values of $f_j$ for disk- and spheroid-dominated galaxies are also found in simulations of galaxy formation \citep{Genel2015, Pedrosa2015, Teklu2015, Zavala2016, Sokolowska2017, El-Badry2018, Rodriguez-Gomez2022}. One of the key challenges in theoretical studies of galaxy formation is to provide a compelling physical explanation for this large difference in angular momentum retention. In effect, this would also serve as an explanation for the main morphological characteristics of galaxies embodied in the Hubble classification scheme.

\begin{acknowledgments}

We thank Min Du for interesting correspondence on this topic and Shy Genel for collaboration on the earlier project from which this one developed.

\end{acknowledgments}

%


\bibliography{ms}{}
\bibliographystyle{aasjournal}

\end{document}